\documentclass[12pt,oneside,notitlepage,abstracton,a4paper]{scrartcl}
\usepackage{epsfig,scrpage2,graphicx}
\usepackage{amsmath}

\renewcommand{\headfont}{\normalfont}

\title{\vspace*{-1.8cm}
\begin{flushright}
{\bf\normalsize LAL/RT 08-01}\\
\vspace*{-0.5cm}
{\bf\normalsize EUROTeV-Report-2007-014}\\
\vspace*{-0.5cm}
{\normalsize February 2008}
\end{flushright}
\vspace*{3cm}
\Large\bf Photon production at the interaction point\\ of the ILC} 
\author{\large\bf R. Appleby\\
{\normalsize\it The Cockcroft Institute and the University of Manchester,\vspace*{-0,3cm} } \\
{\normalsize\it Oxford Road, Manchester, M13 9PL, UK\vspace*{0,5cm}},\\
{\large\bf P. Bambade}\\ {\normalsize\it LAL, Univ Paris-Sud, CNRS/IN2P3, Orsay, France}}

\date{\normalsize }

\begin{document}

\maketitle

\vspace*{0,5cm}
\begin{abstract}
The intense beam-beam effect at the interaction point of the International Linear Collider (ILC) causes large disruption of the beams
and the production of photons. These photons, arising dominantly through beamstrahlung emission, are problematic for
the machine design as they need to be transported and dumped in a controlled way. In this work, we perform simulations of 
the beam-beam interaction to predict photon production rates and distributions for the different beam parameters considered at ILC. 
The results are expressed in terms of a set of cones of excluded power, allowing to define the beam-stay-clear requirements 
relevant for different cases and contexts. A comparison is also made with theoretical expectations. The suggested photon cone half-opening angles are 0.75 and 0.85 mrad in the horizontal and vertical planes, respectively. These cones cover all machine energies and parameter sets, and include the low power Compton photons. 
\end{abstract}

\newpage

\section{Introduction}

%\linenumbers

The International Linear Collider (ILC)~\cite{ilc} is an electron-positron collider with a centre-of-mass energy up to 1 TeV and a 
luminosity goal of $10^{34} \mathrm{cm}^{-2} \mathrm{s}^{-1}$.  The achievement of this goal is challenging, and requires 
nanometre-sized colliding beams at the interaction point, achieved through emittance reductions in dedicated damping rings 
and strong focusing before the interaction point. This, combined with the high bunch charge leads to intense electromagnetic interactions when the beams collide, causing particles of each bunch to be deflected by the space charge of the other. Such deflections
strongly enhance the angular divergence of the outgoing beams and significantly dilute their energy distributions, through radiation of so-called beamstrahlung photons. Re-interactions involving these photons also produce additional secondary charged particles.

The production of beamstrahlung, whilst exploitable as a diagnostic, is generally troublesome to the design of the interaction region and subsequent extraction line. The photons are produced in the direction of the outgoing beam, with considerable power (in the order of MWs), 
in a cone of some opening angle in the forward direction (typically up to a mrad), and must be dumped in a controlled way. They may irradiate parts of the machine, potentially quench superconducting magnets and produce backgrounds in the interaction region detectors, through backscattering. Hence a detailed knowledge and characterisation of the beamstrahlung production is critical to successful linear collider design.  

A further contribution to the photon spectrum arises from the so-called Compton photons. These correspond to quantum interactions between electrons and positrons of the colliding beams, where the accompanying photon radiation has high energy. The dominant
contribution is from the radiative Bhabha process, in the particular kinematic domain involving the exchange of a virtual photon only slightly off mass-shell.  The hard part of this process can be described as Compton scattering of a beam particle on this quasi-real photon, and mostly factorises from the part corresponding to its emission off the electron or positron on the other side. The total power corresponding to these additional photons is however only a few tens of Watts and they are radiated with angles comparable to those of the beamstrahlung photons.

In this report, comprehensive calculations of the beamstrahlung and Compton photon production for the newest ILC beam parameter sets are presented. The resulting photon angular distributions are obtained and a set of cone half-opening angles to represent successive total beamstrahlung powers are presented for each ILC parameter set. In Section~\ref{seccalcs} the computational methods and the ILC parameter sets under consideration are discussed. In Section~\ref{sectheory} approximate analytical expectations of the photon cones are calculated. They are compared to computational results in Section~\ref{secresults}. Conclusions are presented in Section~\ref{secconc}, including the practical beam stay-clear requirements in terms of cone half-opening angles.

\section{ILC parameter space and computational methods}

\label{seccalcs}

In this section, the current ILC beam parameter sets, the key parameters relevant to beamstrahlung production and the method used to simulate the beamstrahlung and Compton photons are described.

The latest ILC beam parameter sets~\cite{ilcparas} are similar to the previous working space for the 500 GeV machine whilst reducing the magnitude of the beam-beam interaction for the 1 TeV case, resulting in smaller mean particle energy losses to beamstrahlung photons, $\delta_B$, at that energy. 

The parameters for the new working space relevant to this work are shown in Table~\ref{tabparasets}, where E, N$_b$, N, $\mathcal{L}$, P and $\sigma_{x,y,z}$ are the centre-of-mass energy, the number of bunches per train,  the number of particles per bunch, the luminosity, the total power contained in the beamstrahlung photons and the r.m.s bunch sizes in the horizontal, vertical and longitudinal dimensions, respectively.

\begin{table}[htb]
\vspace{-0mm}
\begin{center}
\caption{\it Key parameters of the new beam parameter sets used in this work, at 500 GeV and 1 TeV in the centre-of-mass.}
\vspace{0mm}
\begin{tabular}{|l|c|c|c|c|c|c|c|}
\hline
Set		& Energy  & N$_b$ & N [10$^{\mathrm{10}}$] & $\delta_B$ & $\mathcal{L}$ [m$^{-2}$ s$^{-1}$] & P [MW] & $\sigma_{\mathrm{x,y,z}}$\\
&&&&&&& [nm,nm,mm]
\\ \hline \hline
Nominal 	& 500 GeV & 2625 & 2.05 & 0.024 & 2.06E38 & 0.258 & 639, 5.7, 0.3 \\ \hline	
Low Q		& 500 GeV & 5120 & 1.05 & 0.017 & 2.00E38 & 0.184 & 474, 3.5, 0.2 \\ \hline
Large Y		& 500 GeV & 2625 & 2.05 & 0.027 & 2.03E38 & 0.287 & 474, 9.9, 0.5 \\ \hline
Low Power	& 500 GeV & 1320 & 2.05 & 0.055 & 1.99E38 & 0.299 & 474, 3.8, 0.2 \\ \hline
High Lum.	& 500 GeV & 2625 & 2.05 & 0.067 & 4.57E38 & 0.726 & 474, 3.5, 0.15\\ \hline
Nominal 	& 1 TeV   & 2625 & 2.05 & 0.052 & 2.77E38 & 0.897 & 554, 3.5, 0.3 \\ \hline	
Low Q		& 1 TeV	  & 5120 & 1.05 & 0.042  & 2.79E38 & 0.728 & 378, 2.5, 0.2 \\ \hline
Large Y		& 1 TeV	  & 2625 & 2.05 & 0.063 & 2.08E38 & 1.080 & 367, 7.0, 0.6 \\ \hline 
Low Power	& 1 TeV   & 1640 & 2.05 & 0.094 & 2.80E38 & 1.010 & 452, 2.7, 0.2 \\ \hline
High Lum.	& 1 TeV   & 2625 & 2.05 & 0.094 & 4.89E38 & 1.620 & 452, 2.5, 0.2 \\ \hline
\end{tabular}
\label{tabparasets}
\end{center}
\end{table}

The calculation of the beamstrahlung is performed using the beam-beam simulator GUINEA-PIG~\cite{gp}. This code models the 
interaction of two colliding flat Gaussian beams in a consistent way, by calculating particle trajectories from Lorentz's force law and recomputing the relevant electromagnetic fields as the bunches traverse eachother. Beamstrahlung emission and secondary particles produced during beam collisions, e.g. Compton photons, can also be generated. The simulated data used in this work are available at~\cite{beamfiles}.

Resulting photon distributions are presented in terms of cones with half-opening angles corresponding to a given amount of total excluded photon power. This definition~\cite{beamcones2004} is attractive because it can be directly related to the beam stay-clear requirements of the beamline, for each case in consideration of its specific tolerance (e.g. for quench limits or irradiation). It is moreover less sensitive to statistical fluctuation and hence can describe more correctly the impact of the far tails of the photon distribution. An alternative definition, expressed in terms of a maximum production angle and which is often used in the ILC community, can be recovered by tending the excluded power to zero. 

In the case of primary beamstrahlung photon production, the maximum size of the horizontal cone opening angle occurs for zero-offset bunch collisions. However the electromagnetic field distribution in the vertical plane results in a maximum vertical cone opening angle at some vertical offset between the beams at the interaction point, such that the beamstrahlung parameter~$\delta_B$ is maximised. In the vertical plane, cone half-opening angles were therefore computed after performing vertical scans to determine this (parameter-set dependent) worst-case offset~\cite{beamcones2004}. In the case of Compton photon production, the rate of Compton electrons is on the other hand essentially proportional to the instantaneous luminosity and so the flux of Compton photons falls rapidly with beam offsets in both planes. Hence the worst-case for Compton photon production in both planes is for zero beam offset at the collision point.

\section{Theoretical Expectations}

\label{sectheory}

The beamstrahlung  photon angles produced by the beam-beam simulation can be compared  to theoretical expectations. The rigid-beam model of two colliding bunches developed in~\cite{bb1} makes the assumption of a Gaussian beam charge distribution for the three
coordinates (x, y, z): 
\begin{equation}
\rho(x,y,x)=\frac{N e}{(2\pi)^{3/2}\sigma_x \sigma_y \sigma_z} 
\exp\left(-\frac{x^2}{2\sigma_x^2}-\frac{y^2}{2\sigma_y^2}-\frac{s^2}{2\sigma_z^2}\right),
\end{equation}
where e is the electron charge, which remains unchanged as the relativistic bunches collide. In this model, the pinch effect is hence neglected, an approximation which is not justified in the vertical plane at the ILC. The equations are valid for arbitrary transverse beam sizes, and in general $\sigma_x \neq \sigma_y$. The calculation of the beam-beam kicks proceeds by computation of the fields of the bunch, followed by calculation of the motion of particles in the other bunch using the Lorentz's force law.
  
Individual beam-beam kicks arise from the interaction of single beam particles with the opposite bunch's field. For a beam particle located at position $(x,y)$, the following expression can be used~\cite{bb1}:
\begin{eqnarray}
\Delta x'-\imath \Delta y'=\frac{2 N r_e}{\gamma} \frac{\imath \sqrt{\pi}}{\sqrt{2 (\sigma_x^2-\sigma_y^2)}}
\large(
\mathrm{w}\left(\frac{x+\imath y}{\sqrt{2(\sigma_x^2-\sigma_y^2)}}\right)\nonumber \\
-\exp\left(-\frac{x^2}{2\sigma_x^2}-\frac{y^2}{2\sigma_y^2}\right)\mathrm{w}\left(\frac{x/R+\imath yR}{\sqrt{2(\sigma_x^2-\sigma_y^2)}}\right)\large),
\label{equinco}
\end{eqnarray}
where $r_e$ is the classical electron radius and $R=\sigma_x/\sigma_y$. The complex error function is denoted by $\mathrm{w}(z)$ and is defined as:
\begin{equation}
\mathrm{w}(z)=\exp(-z^2)\mathrm{Erfc}(\imath z),
\end{equation}
where $\mathrm{Erfc}(z)$ denotes the complementary error function, defined by $\mathrm{Erfc}(z)=1-\mathrm{Erf}(z)$. 

The overall kick of the entire bunch in the field of the other bunch is simply the average of the individual kicks of its particles. This kick is described in terms of the overall bunch offsets in the horizontal and vertical directions and is also zero for perfectly aligned bunches. The expression which is obtained is the same as for the individual kick, with the substitutions $(x,y)\rightarrow(X,Y)$, where
$X$ and $Y$ denote the horizontal and vertical bunch offsets, and $2\sigma_{x,y}^2\rightarrow4\sigma_{x,y}^2$:
\begin{align}
\Delta x'-\imath \Delta y'=\frac{2 N r_e}{\gamma} \frac{\imath \sqrt{\pi}}{\sqrt{4 (\sigma_x^2-\sigma_y^2)}}
\huge(
\mathrm{w}\left(\frac{X+\imath Y}{\sqrt{4(\sigma_x^2-\sigma_y^2)}}\right)\nonumber \\
-\exp\left(-\frac{X^2}{4\sigma_x^2}-\frac{Y^2}{4\sigma_y^2}\right)\mathrm{w}\left(\frac{X/R+\imath YR}{\sqrt{4(\sigma_x^2-\sigma_y^2)}}\right)
\huge).
\end{align}

Because of the averaging, the maximum individual kick always exceeds the overall one. Since the radiated beamstrahlung photons are strongly peaked in the forward direction of the emitting particle, the maximum prodution angle for photons equals that for individual particles after adding the natural beam divergence of the incoming beams prior to colliding, $\sigma_x'=\sqrt{\epsilon_x/\beta_x^*}$.

\begin{figure*}[tb]
\vspace{0mm} \centering
\includegraphics[width=90mm]{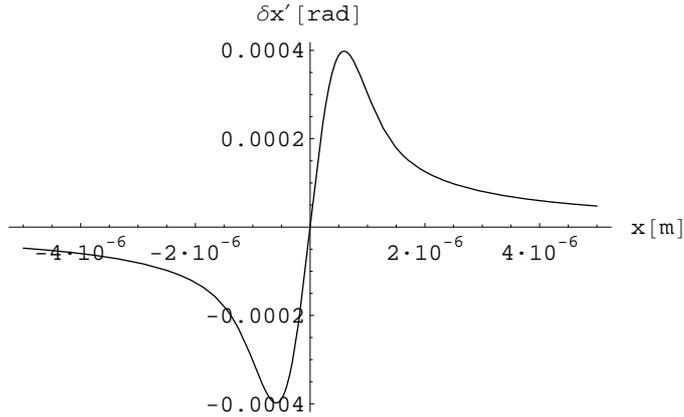}
\vspace{0mm} \caption{\it The individual particle kick for the 1 TeV Low Power beam parameter set.}
\label{figincoherentkick}
\end{figure*}

\begin{figure*}[tb]
\vspace{5mm} \centering
\includegraphics[width=90mm]{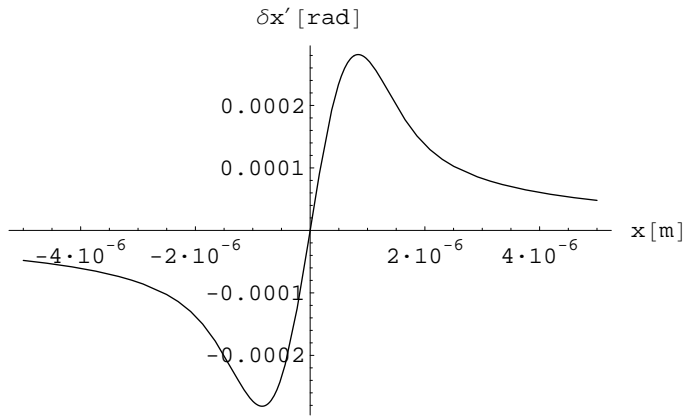}
\vspace{0mm} \caption{\it The overall bunch kick for the 1 TeV Low Power beam parameter set.}
\label{figcoherentkick}
\vspace*{0,6cm}
\end{figure*}

Numerical results for the individual kicks are shown in Table~\ref{tabkicks} for the beam parameter sets with 
the largest beam-beam effects. 
The individual and overall kicks are also shown as a function of horizontal beam offsets in Figures~\ref{figincoherentkick} and~\ref{figcoherentkick} for the Low Power beam parameter set at 1 TeV. 

\begin{table}[htb]
\vspace{-0mm}
\begin{center}
\caption{\it Photon production angles for the new ILC beam parameter sets~\cite{ilcparas}.}
\vspace{2mm}
\begin{tabular}{|c|c|c|c|c|c|}
\hline
$E_{\mathrm{CM}}$ & Parameter set & $\sigma_x$ [nm]  & $\sigma_x'$ [mrad] & Kick maximum [mrad] & $3\sigma_x'$+kick \\ \hline \hline
500 & Nominal & 639 & 0.032 & 0.280 & 0.376 \\ \hline
500 & Large Y & 474 & 0.043 & 0.379 & 0.508 \\ \hline
500 & Low Power & 474 & 0.043 & 0.378 & 0.508 \\ \hline 
500 & High Lumi & 474 & 0.043 & 0.378 & 0.508 \\ \hline
1 TeV & Nominal & 554 & 0.018 & 0.324 & 0.379 \\ \hline
1 TeV & Large Y & 367 & 0.033 & 0.489 & 0.589 \\ \hline
1 TeV & Low Power & 452 & 0.023 & 0.397 & 0.465 \\ \hline
1 TeV & High Lumi & 452 & 0.023 & 0.397 & 0.465 \\ \hline
\end{tabular}
\label{tabkicks}
\end{center}
\end{table}

\section{Results}

\label{secresults}

\begin{figure}[h]
\begin{center}
\begin{minipage}{0.48\textwidth}
\includegraphics[width=7.7cm]{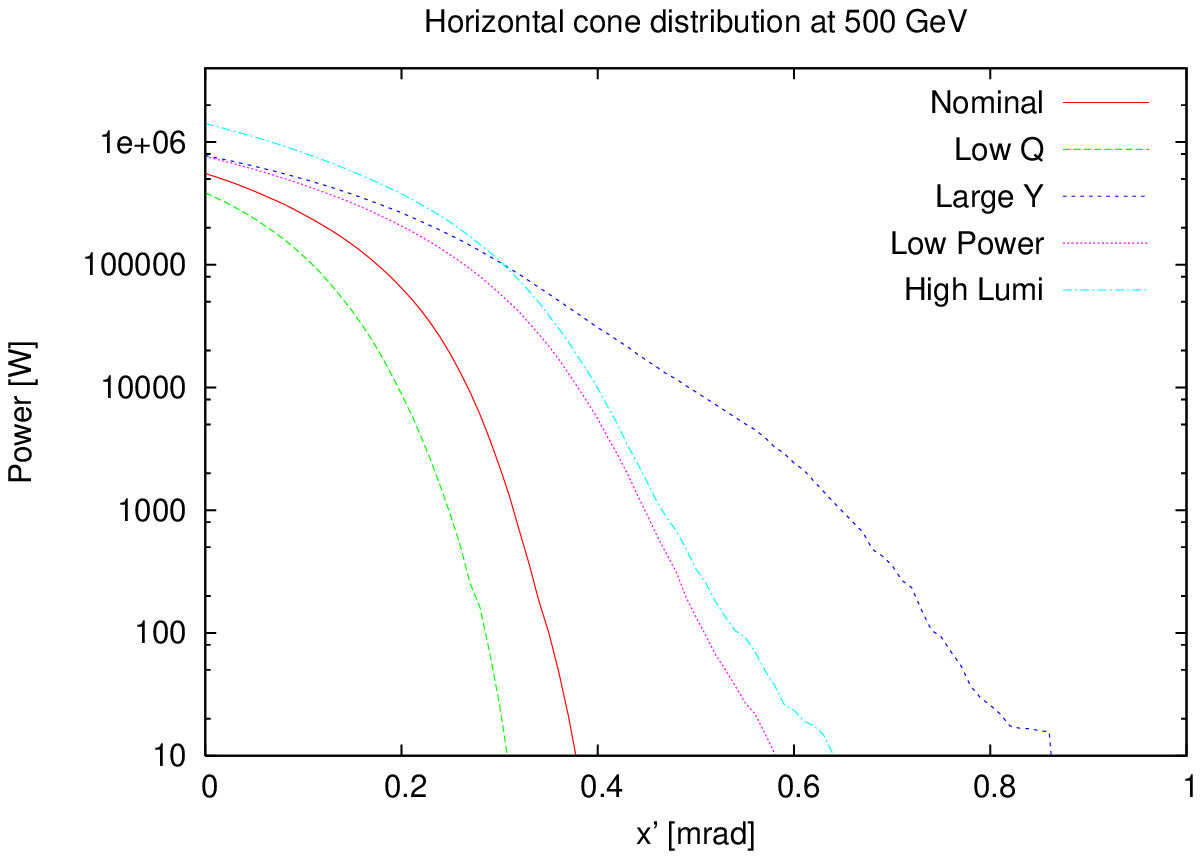}
\end{minipage}
\hfill
\begin{minipage}{0.48\textwidth}
\includegraphics[width=7.7cm]{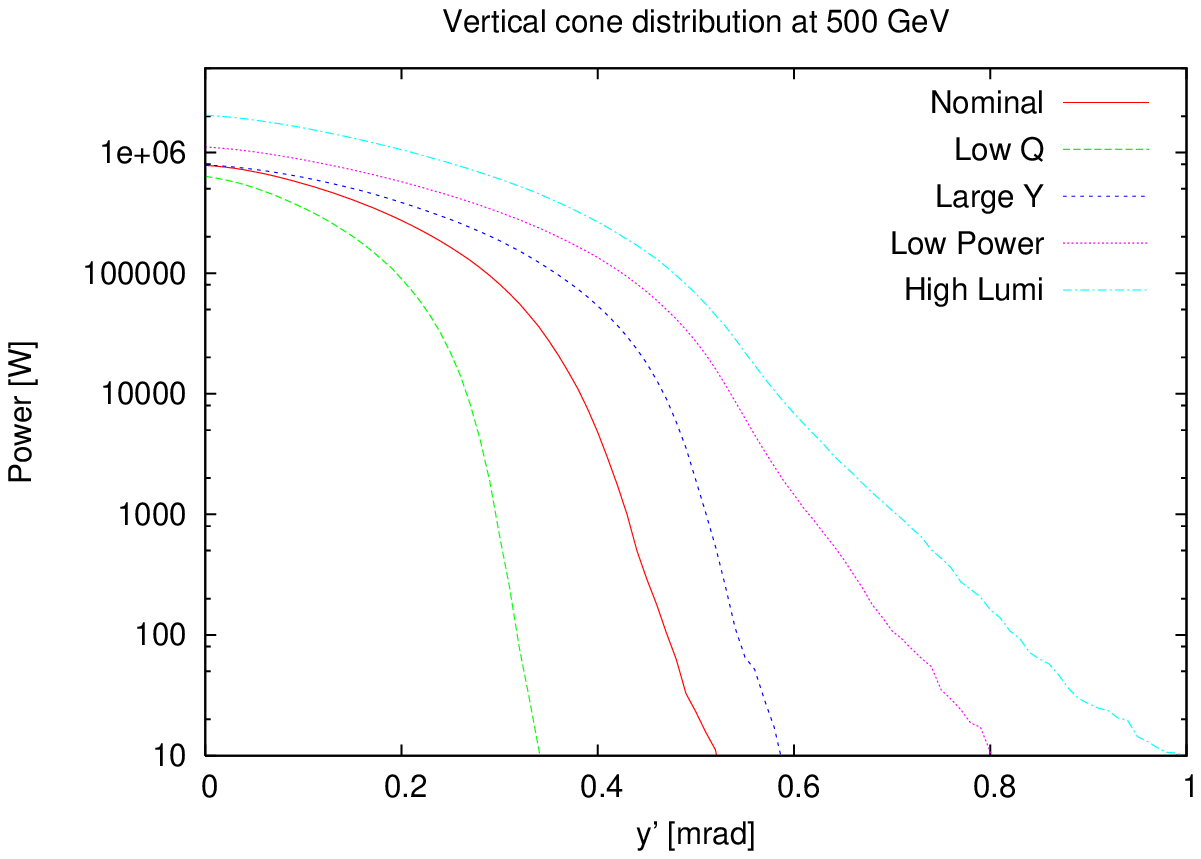}
\end{minipage}
\caption{\it The primary beamstrahlung photon distribution for the 500 GeV machine in the horizontal plane (left) and the vertical plane (right).}
\label{fig500prim}
\end{center}
\end{figure}

\begin{figure}[h]
\begin{center}
\begin{minipage}{0.48\textwidth}
\includegraphics[width=7.7cm]{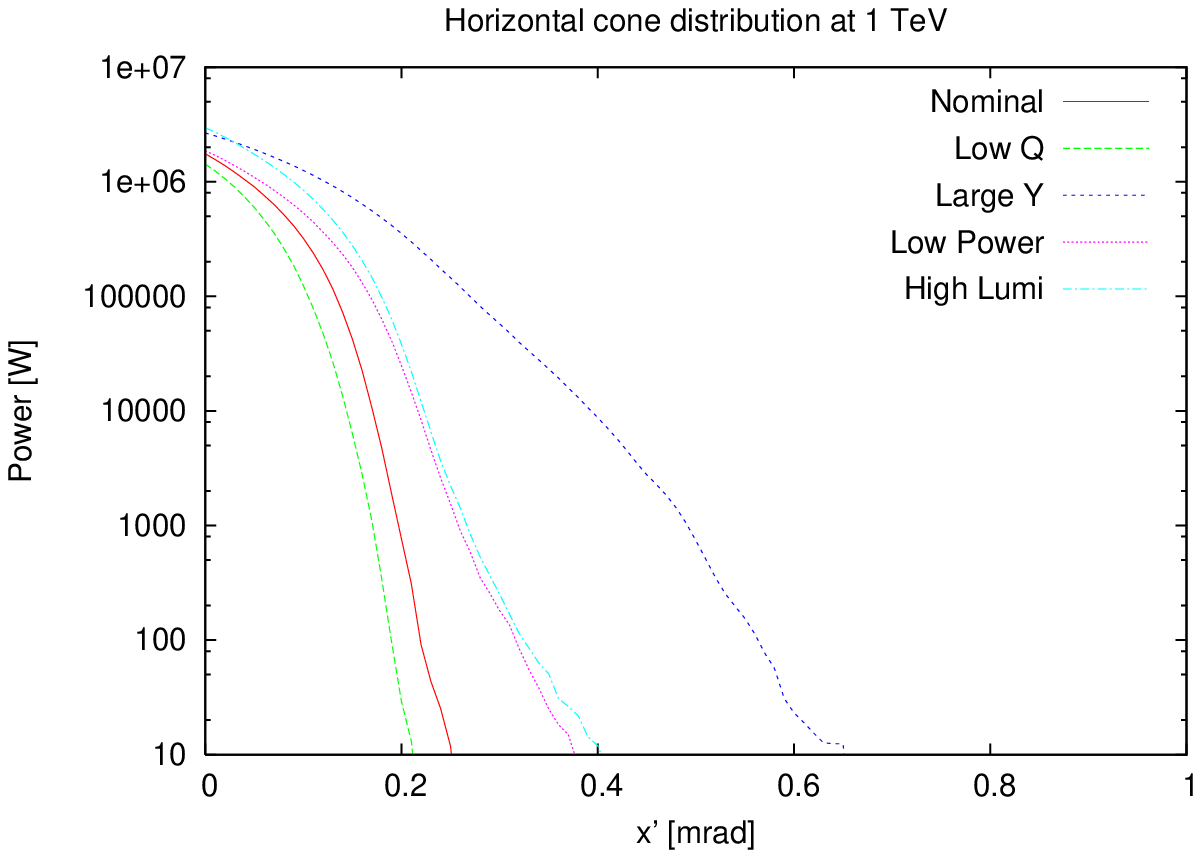}
\end{minipage}
\hfill
\begin{minipage}{0.48\textwidth}
\includegraphics[width=7.7cm]{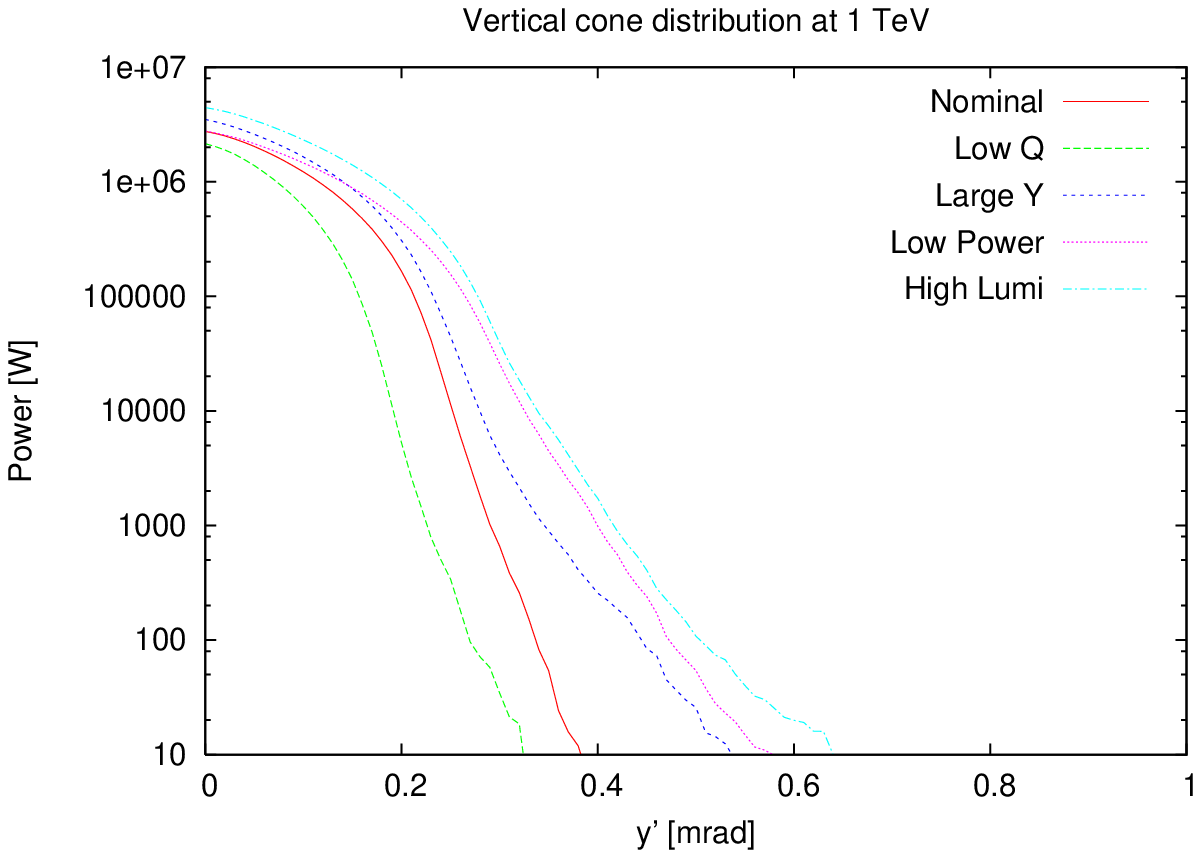}
\end{minipage}
\caption{\it The primary beamstrahlung photon distribution for the 1 TeV machine in the horizontal plane (left) and the vertical plane (right).}
\label{figtevprim}
\end{center}
\end{figure}

Cones of excluded photon power calculated using GUINEA-PIG are presented in this section. For the horizontal plane, zero-offset colliding beams are used, while for the vertical plane, the offset maximising the beamstrahlung power is determined by scanning. The natural beam divergences of the incoming beams prior to colliding are taken into account automatically by this calculation, however the effect is minor since the corresponding standard deviations are much smaller than individual deflection angles and since they enter in quadrature in the excluded power computations. Figure~\ref{fig500prim} shows the distributions obtained for the five beam parameter sets at 500 GeV in the centre-of-mass. The variation with beam parameters can be clearly seen, and sets with large energy losses to beamstrahlung and/or large pinch effects show the largest photon cone opening angles. The 1 TeV machine horizontal and vertical distributions are shown in Figure~\ref{figtevprim}.

Tables~\ref{tab500beamcone} and~\ref{tabtevbeamcone} give the numerical values of the cone half-opening angles for the 500 GeV and 1 TeV machines, respectively, for the beam parameters in Figures~\ref{fig500prim} and~\ref{figtevprim} and corresponding to three successive criteria on excluded power, at 100, 10 and 1 W. Maximum values over the different beam parameters and centre-of-mass energies are summarised in Table~\ref{tabprimangles}. 

Clearly, specifying a given amount of excluded power is context-dependent, and somewhat arbitrary. The 100, 10 and 1 W criteria are 
suggested here as they correspond to typical order-of-magnitude tolerances for either activation of beam-line components (100 W) or quenching in super-conductive magnets (1-10 W)~\cite{2mrad-JINST}. However, in principle any criterion thought to be relevant for a particular application can be specified from the curves in Figures~\ref{fig500prim} and~\ref{figtevprim}. For completeness, cone half-opening angles for 1, 10 and 100 kW criteria are also given in Table~\ref{tabprimanglesextra}.

Comparisons to the theoretical calculations presented in Section~\ref{sectheory} can now be made. 

\begin{table}[h!]
\vspace{-0mm}
\begin{center}
\caption{\it Horizontal and vertical beamstrahlung cone half-opening angles at 500 GeV for the different parameter sets, and with
100 W, 10 W and 1 W definitions for the photon cone.}
\vspace{2mm}
\begin{tabular}{|l|c|c|c|c|}
\hline
Set		& Plane & 100 W & 10 W & 1 W \\ \hline \hline
Nominal [mrad]	& X & 0.35 & 0.37 & 0.39 \\ \hline
Low Q	[mrad] 	& X & 0.28 & 0.30 & 0.34\\ \hline
Large Y [mrad]	& X & 0.74& 0.86 & 0.88 \\ \hline
Low Power [mrad]	& X &0.50 & 0.58&0.70 \\ \hline
High Luminosity [mrad]& X &0.54 &0.64 &0.66 \\ \hline
Nominal [mrad]	& Y &0.47 &0.52 &0.55 \\ \hline
Low Q [mrad]	 	& Y &0.31 &0.34 & 0.35\\ \hline
Large Y [mrad]	& Y &0.54 &0.58 &0.60 \\ \hline
Low Power [mrad]	& Y &0.70 &0.80 &0.88 \\ \hline
High Luminosity [mrad] & Y &0.82 &0.99 & 1.09\\ \hline
\end{tabular}
\label{tab500beamcone}
\end{center}
%\end{table}

%\begin{table}[h]
\vspace{-0mm}
\begin{center}
\caption{\it Horizontal and vertical beamstrahlung cone half-opening angles at 1 TeV for the different parameter sets, and with
100 W, 10 W and 1 W definitions for the photon cone.}
\vspace{2mm}
\begin{tabular}{|l|c|c|c|c|}
\hline
Set		& Plane & 100 W & 10 W & 1 W \\ \hline \hline
Nominal [mrad]	& X &0.21 &0.25 &0.27 \\ \hline
Low Q [mrad]	 	& X &0.18 & 0.21&0.22 \\ \hline
Large Y [mrad] 	& X &0.56 &0.65 & 0.67\\ \hline
Low Power [mrad]	& X &0.31 & 0.37& 0.44\\ \hline
High Luminosity [mrad]& X &0.32 &0.40 & 0.47\\ \hline
Nominal [mrad]	& Y &0.33 &0.38 &0.43 \\ \hline
Low Q [mrad]	 	& Y &0.26 & 0.32&0.38 \\ \hline
Large Y [mrad]	& Y &0.44 & 0.53&0.59 \\ \hline
Low Power [mrad]	& Y &0.47 & 0.57&0.63 \\ \hline
High Luminosity [mrad]& Y &0.50 &0.63 &0.72 \\ \hline
\end{tabular}
\label{tabtevbeamcone}
\end{center}
%\end{table}

%\begin{table}[h]
\vspace{-0mm}
\begin{center}
\caption{\it Horizontal and vertical maximum beamstrahlung cone half-opening angles, covering all parameter sets and both energies, for 1, 10 and 100 W of excluded power.}
\vspace{1mm}
\begin{tabular}{|l|c|c|c|}
\hline
Plane		& 100 W & 10 W & 1 W \\ \hline \hline
Horizontal [mrad]	& 0.74 & 0.86 &0.88 \\ \hline
Vertical [mrad]	& 0.82 & 0.99 &1.09 \\ \hline
\end{tabular}
\label{tabprimangles}
\end{center}
\end{table}

\begin{table}[h]
\vspace{-0mm}
\begin{center}
\caption{\it Horizontal and vertical maximum beamstrahlung cone half-opening angles, covering all parameter sets and both energies, for 1, 10 and 100 kW of excluded power.}
\vspace{4mm}
\begin{tabular}{|l|c|c|c|}
\hline
Plane		& 1 kW & 10 kW & 100 kW \\ \hline \hline
Horizontal [mrad]	& 0.64 & 0.49 & 0.30\\ \hline
Vertical [mrad]	& 0.70 & 0.58 & 0.47\\ \hline
\end{tabular}
\label{tabprimanglesextra}
\end{center}
\end{table}

Since they use the rigid beam model, these calculations are only valid in the horizontal plane and for cases with small disruption. 

Comparing Tables~\ref{tab500beamcone} and~\ref{tabtevbeamcone} indicates that the calculations predict the maximum horizontal angle of photon emissions reasonably well only for some of the parameter sets. For example, the 500 GeV Nominal theoretical maximum angle is 0.376 mrad, which agrees well with the 0.35-0.39 mrad range computed from the beam-beam simulation for the 100 - 1 W criteria. This is however not the case for the Large Y and High Luminosity sets, where some differences are seen. Such differences are attributed to extreme colliding particle trajectories in GUINEA-PIG. These may originate from the input angular distribution of the beams and may also result from some disruption occuring for these parameter sets. In all cases, the GUINEA-PIG calculation gives larger values and should be used for conservative results.

\begin{figure}[h]
\begin{center}
\begin{minipage}{0.48\textwidth}
\includegraphics[width=7.7cm]{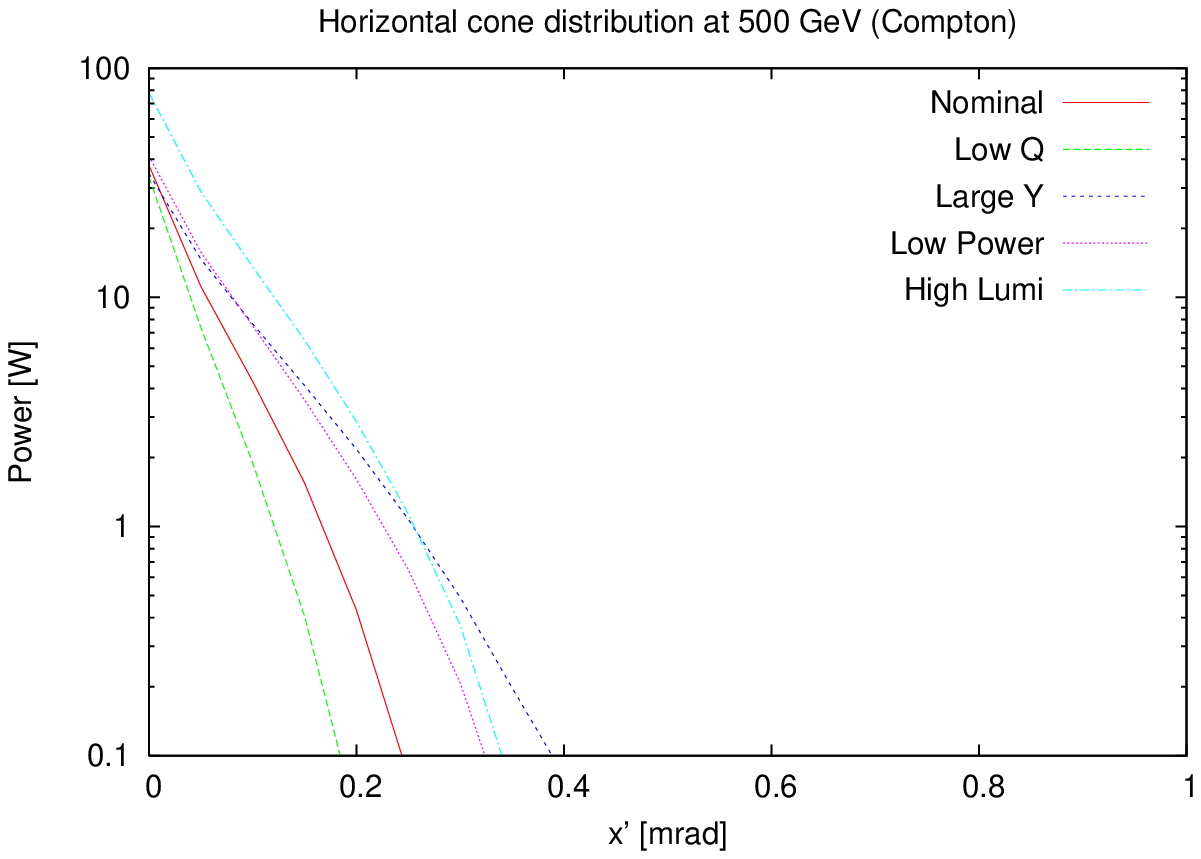}
\end{minipage}
\hfill
\begin{minipage}{0.48\textwidth}
\includegraphics[width=7.7cm]{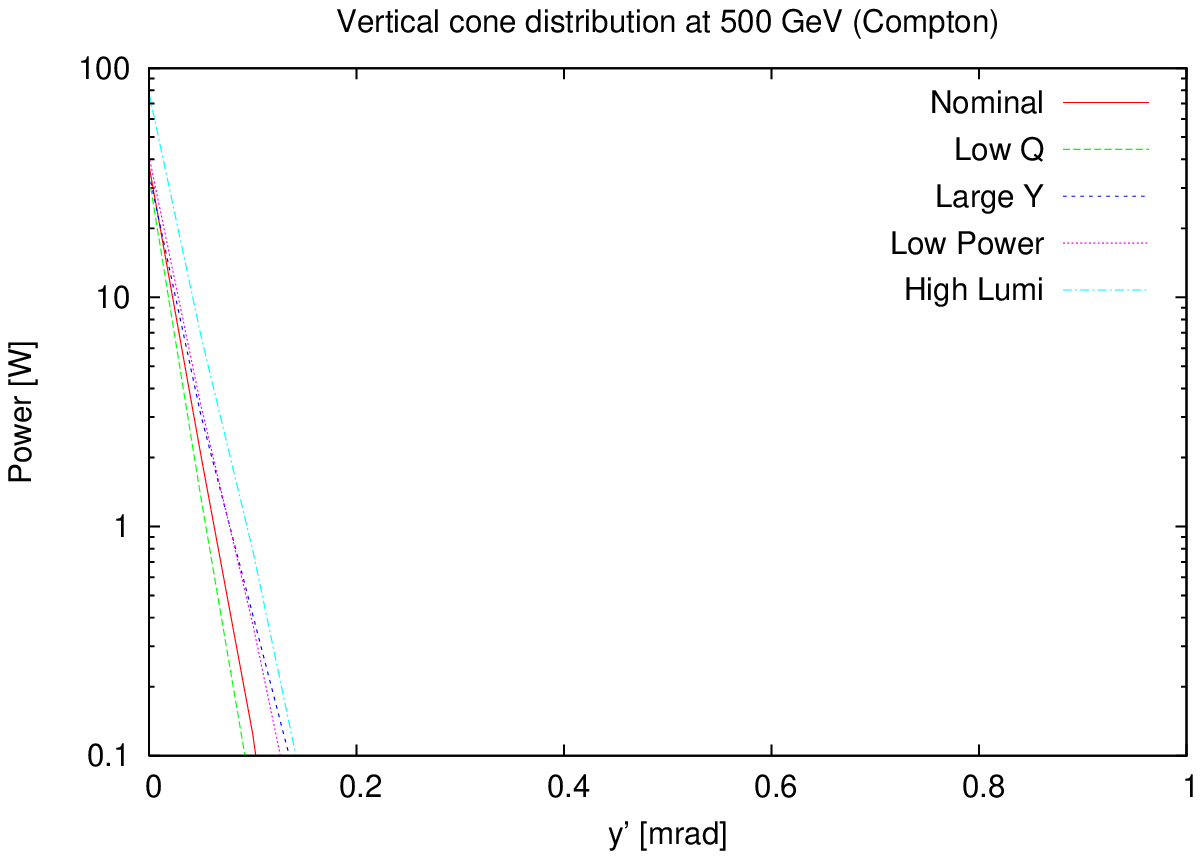}
\end{minipage}
\caption{\it Compton photon distribution for the 500 GeV machine in the horizontal (left) and vertical (right) planes.}
\label{fig500compt}
\end{center}
\end{figure}

\begin{figure}[h]
\begin{center}
\begin{minipage}{0.47\textwidth}
\includegraphics[width=7.5cm]{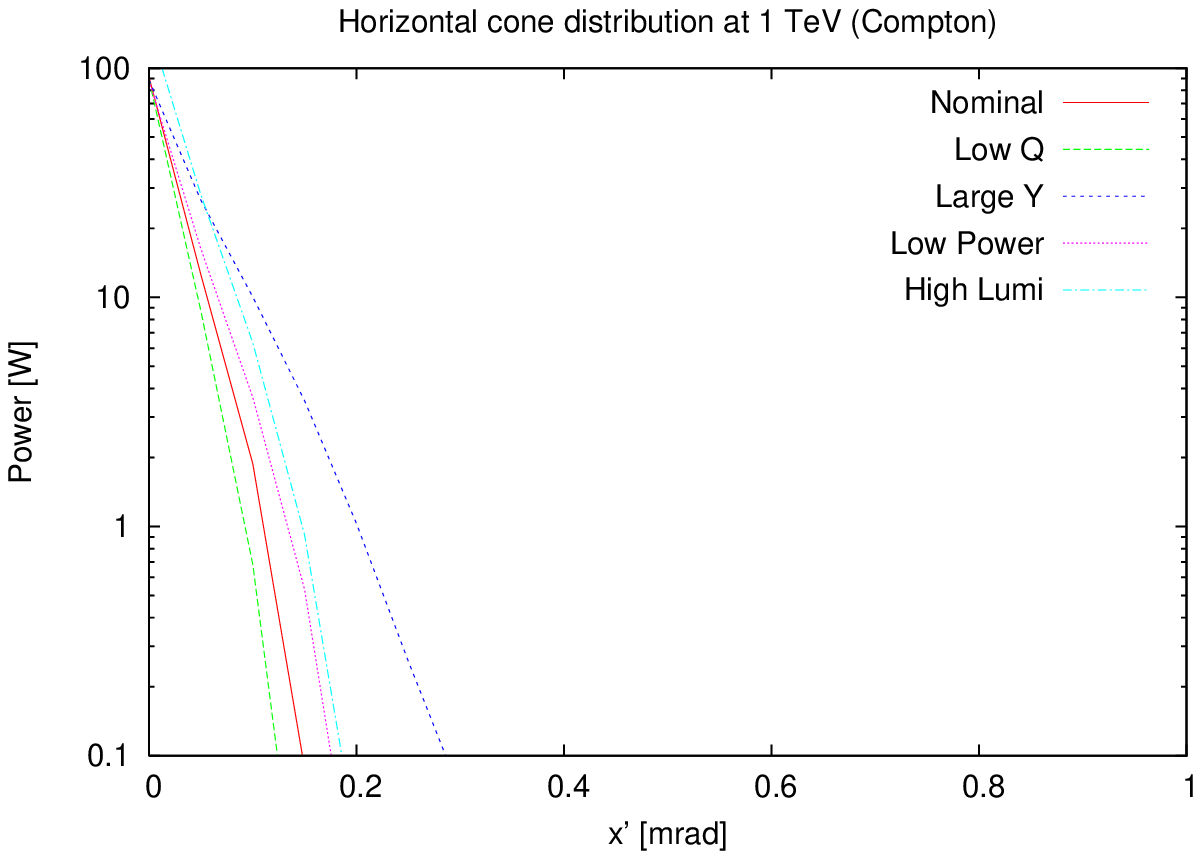}
\end{minipage}
\hfill
\begin{minipage}{0.47\textwidth}
\includegraphics[width=7.5cm]{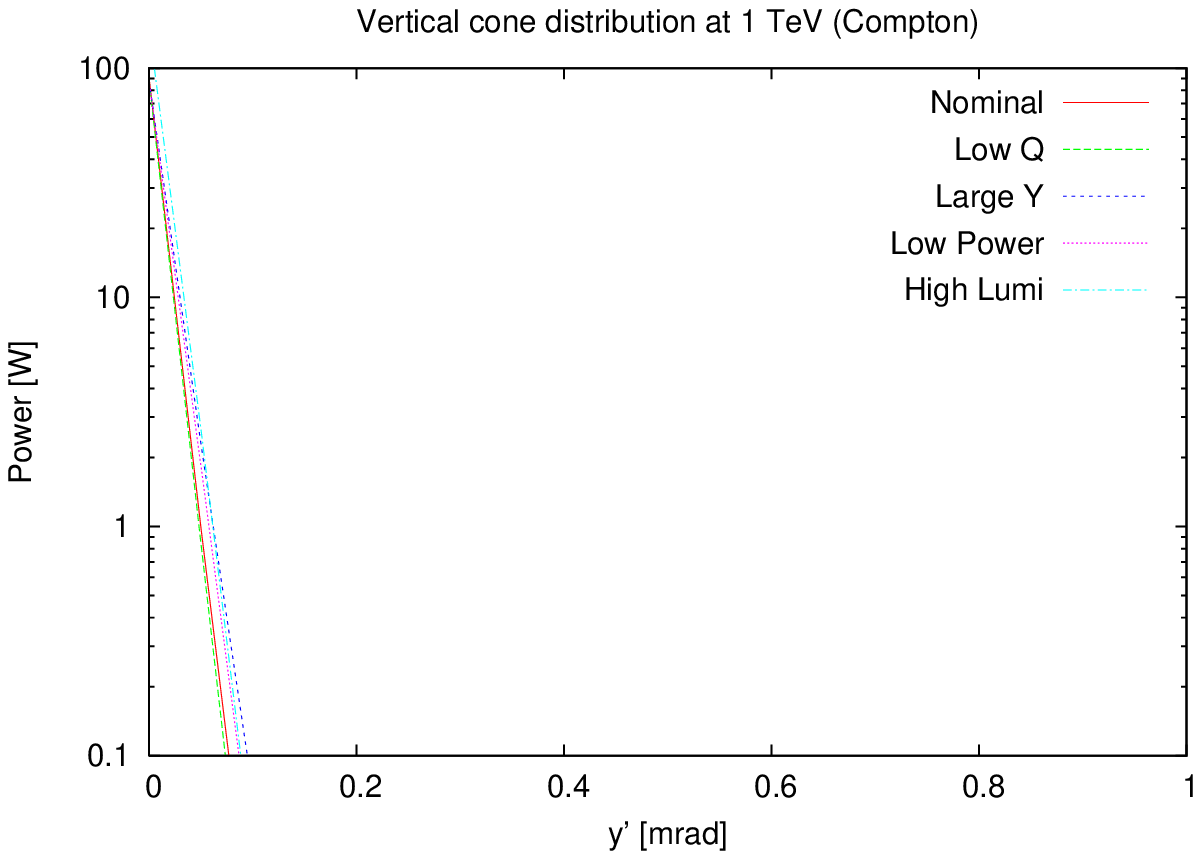}
\end{minipage}
\caption{\it Compton photon distribution for the 1 TeV machine in the horizontal (left) and vertical (right) planes.}
\label{figtevcompt}
\end{center}
\end{figure}

Similar distributions for the Compton photons are presented in Figures~\ref{fig500compt} and~\ref{figtevcompt}, in the horizontal and vertical planes and for the five parameter sets. While extending to similar angles as for beamstrahlung photons, the corresponding power is orders of magnitude less. The maximum cone half-opening angles previously computed from beamstrahlung are hence not modified.

\section{Conclusions}

In this paper, the angles of photons emitted through the beam-beam interaction at the International Linear Collider have been computed using both the rigid beam model and the beam-beam simulator GUINEA-PIG. These calculations, and the subsequent understanding of the beamstrahlung, are crucial to the effective design of the interaction region and extraction line. 

The rigid beam theoretical predictions tend to underestimate the photon cone opening angles for some of the beam parameter sets, particularly for cases with large disruption. GUINEA-PIG is hence used for conservative results.

Cones of excluded photon power are defined corresponding to tolerances relevant to different aspects of beam-line design: 100 W for general activation and 10 - 1 W for super-conductive magnet quenching. Two contributions were studied, from beamstrahlung emission and from a Compton process involving the exchange of a slightly virtual photon between the incident electrons and positrons. By far the first contribution dominates for most of the angles. The results are summarised in Table~\ref{tabprimangles}.

For the general design of the extraction line, the 100 W definition is applicable. In this case, half-opening angles of 0.75 and 0.85 mrad can be used as conservative definitions of the required beam-stay-clear in the horizontal and vertical planes, respectively. 

\label{secconc}

\section*{Acknowledgement\vspace*{-0,3cm}}
We would like to thank Daniel Schulte, Olivier Napoly, Olivier Dadoun, Mark Briscombe, Daniel Smallqy and C{\'e}cile Rimbault for helpful advice and assistance, and special thanks to Deepa Angal-Kalinin for checking the manuscript and useful advice.
This work was supported by the Commission of the European Communities under the 6th Framework Programme "Structuring
the European Research Area", contract number RIDS-011899, and by the Alliance programme of the British Council and Minist{\`e}re des Affaires {\'E}trang{\`e}res.

\end{document}